\documentclass[english,aps,prd,superscriptaddress,notitlepage,nofootinbib,preprintnumbers,floatfix]{revtex4-1}
\usepackage[T1]{fontenc} \usepackage[latin9]{inputenc}
 \usepackage{babel} \usepackage{array}
\usepackage{float} \usepackage{multirow} \usepackage{amsmath}
\usepackage{graphicx} \usepackage{esint} \usepackage[unicode=true,
  bookmarks=false, breaklinks=false,pdfborder={0 0
    1},backref=section,colorlinks=false] {hyperref} \hypersetup{
  colorlinks}

\makeatletter

\providecommand{\tabularnewline}{\\}

\def\Mpl{M_{\rm P}}

\usepackage{babel}

\makeatother

\begin{document}
\preprint{YITP-20-65}
\title{Reducing the $H_{0}$ tension with generalized Proca theory}
\author{Antonio De~Felice} \email{antonio.defelice@yukawa.kyoto-u.ac.jp}
\affiliation{Center for Gravitational Physics, Yukawa Institute for
  Theoretical Physics, Kyoto University, 606-8502, Kyoto, Japan}

\author{Chao-Qiang Geng} \email{geng@phys.nthu.edu.tw}
\affiliation{Department of Physics, National Tsing Hua University,
  Hsinchu, Taiwan 300} \affiliation{National Center for Theoretical
  Sciences, Hsinchu, Taiwan 300} \affiliation{School of Fundamental
  Physics and Mathematical Sciences, Hangzhou Institute for Advanced
  Study, UCAS, Hangzhou 310024, China} \affiliation{International
  Centre for Theoretical Physics Asia-Pacific, Beijing/Hangzhou,
  China}

\author{Masroor C.~Pookkillath} \email{masroor.cp@yukawa.kyoto-u.ac.jp}
\affiliation{Center for Gravitational Physics, Yukawa Institute for
  Theoretical Physics, Kyoto University, 606-8502, Kyoto, Japan}

\author{Lu Yin} \email{yinlu@gapp.nthu.edu.tw}
\affiliation{Department of Physics, National Tsing Hua University,
  Hsinchu, Taiwan 300} \affiliation{School of Fundamental Physics and
  Mathematical Sciences, Hangzhou Institute for Advanced Study, UCAS,
  Hangzhou 310024, China} \affiliation{International Centre for
  Theoretical Physics Asia-Pacific, Beijing/Hangzhou, China}

\begin{abstract}
We investigate the cosmological viability of the generalized proca
theory. We first implement the background and linear perturbation
equations of motion in the Boltzmann code and then study the
constraints on the parameters of the generalized proca theory after
running MCMC against the cosmological data set. With Planck + HST
data, we obtain the constraint $h=0.7334_{-0.0269}^{+0.0246}$, which
indicates that the tension between early universe and late time
universe within this theory is removed. By adding other late-time data
sets (BAO, RSD, etc.) we show that the tension is reduced, as the
2$\sigma$ allowed region for $h$ in Proca,
$h=0.7041_{-0.0087}^{+0.0094}$, overlaps with the 2$\sigma$ region of
the HST data.
\end{abstract}
\maketitle

\section{Introduction}

The cosmological parameters ($H_{0}$, etc.) characterize our universe
and explain how our universe evolves during its various stages. Out of
these parameters some are measured from the background evolution and
others are measured from the linear perturbation theory. Although the
measurements of these parameters have become very precise, these show
up some tensions in the expansion rate of the universe today,
$H_{0}$~\cite{Bernal:2016gxb}. This tension -- if one makes the strong
prior that the theoretical model we have, i.e.\ $\Lambda$CDM, is
complete -- is considered to be due to unknown systematics in the
early or late-time universe measurements of
$H_{0}$~\cite{Verde:2019ivm}.  In the last year, there were around six
new independent methods for the estimation of $H_{0}$. All these
indicate that the tension does not depend on any methodology being
used in the measurement~\cite{Verde:2019ivm}, as such, it points more
and more towards an embarrassing and puzzling picture of the
cosmological background evolution.

As far as the early universe measurements are concerned the most
important estimation of the cosmological parameters comes from the
Cosmic Microwave Background (CMB)
radiation~\cite{Aghanim:2018eyx}. The probe is made by the Planck
Collaboration and represents one of the most significant and precise
measurement in the context of cosmology. Deducing $H_{0}$ from CMB can
be considered as a process of three steps. First, the determination of
baryon and matter densities to calculate the comoving sound horizon at
the last scattering epoch, $r$. Second, infer the angular size of the
last scattering surface, $\theta$ from the spacing between acoustic
peak to find the comoving angular diameter distance to the last
scattering surface, $D=r/\theta$. Finally, the relation
$D=\int_{0}^{z}dz/H\left(z\right)$, although evaluated at high
redshifts ($z\simeq1060$), still depends on the dynamical history of
$H(z)$, so that, given a model, one can infer the value taken by
$H(z)$ today.  In the first step, the determination of baryon and
matter densities we need to assume a theory (the Planck Collaboration
considers $\Lambda\text{CDM}$ as the
theory~\cite{Aghanim:2016sns}). The Planck Collaboration measures the
expansion rate today to be,
$H_{0}=67.4{}_{-0.5}^{+0.5}\text{\ km\,s\ensuremath{^{-1}}\,Mpc}^{-1}$
with a remarkable precision of $1\%$~\cite{Aghanim:2018eyx}. Another
independent measurement from the early universe data
is~\cite{Abbott:2017smn}, which predicts an expansion rate today
$H_{0}=67.4_{-1.2}^{+1.1}\text{\ km\,s\ensuremath{^{-1}}\,Mpc}^{-1}$
combining observation DES+BAO+BBN. The value of $H_{0}$ from the early
universe data obtained with $\Lambda\text{CDM}$ as a prior is
significantly smaller than the one measured from the late time data,
which is named as a direct measurement.

From the late time universe data the most prominent measurement of
$H_{0}$ is from the Hubble Space Telescope (HST)~\cite{Riess:2019cxk}.
This experiment observes the peak brightness of type Ia supernova,
which can be used as a distance ladder. Type Ia supernova is
calibrated with Cepheid Period - Luminosity relation, which is in the
Large Magellanic Cloud. This gives an excellent opportunity to
determine $H_{0}$ without assuming any theory. With the improved
measurements and calibrations, HST measures the expansion rate today
as
$H_{0}=74.03_{-1.42}^{+1.42}\text{\ km}\text{s}^{-1}\text{Mpc}^{-1}$\cite{Riess:2019cxk}.

Apart from HST measurement, recently there have been different
techniques to measure Hubble expansion using late time observational
data, viz.  $\text{H}0\text{LiCOW}$~\cite{Wong:2019kwg}, Megamaser
Cosmology Project (MCP)~\cite{Reid:2008nm}, Carnegie-Chicago Hubble
Program (CCHP) Collaboration~\cite{Freedman:2019jwv},
etc. $\text{H}0\text{LiCOW}$ exploits strong gravitational lensing to
measure the quasar system and uses flat $\Lambda\text{CDM}$ to measure
$H_{0}$, obtaining
$H_{0}=73.3_{-1.8}^{+1.7}\text{\ km}\text{s}^{-1}\text{Mpc}^{-1}$\cite{Wong:2019kwg},
which is in agreement with HST. The other late time measurements are
also in close agreement with the SH$0$ES Collaboration. All these
indicate that there is a strong disagreement in the prediction of
$H_{0}$ from the early universe data using $\Lambda\text{CDM}$ and
late time universe between $4.0\sigma$ and
$5.8\sigma$~\cite{Verde:2019ivm}.  As stated above this tension does
not depend on the methodology. This opens a room to explore for
theoretical ideas to address this tension.

There are several approaches to address this growing tension in
cosmology.  In general it can be classified as pre -- recombination
solution and post -- recombination solution, where the recombination
occurred at the redshift
$z\simeq1100$~\cite{Verde:2019ivm,Knox:2019rjx}.  There has been a
study on $H_{0}$ tension which assumes a scalar field which acts as an
early dark energy at the redshift $z\gtrsim3000$ and it decays like
radiation~\cite{Poulin:2018cxd}. We approach this tension from the
point of view of modified gravity, which can be classified into a post
recombination solution.

In modified gravity scenarios, especially in the context of late time
modified gravity, there is, in many cases, a single extra degree of
freedom, which is responsible for the universe to
accelerate~\cite{Nojiri:2017ncd}.  Generally, these theories have an
effective equation of state for dark energy $w_{\text{DE}}$, which
takes different values at different redshifts, e.g.\ for scalar tensor
theories~\cite{Fujii:2003pa}, vector tensor
theories~\cite{Heisenberg:2014rta,Heisenberg:2017mzp}, etc. In this
paper we will consider one of the simplest Generalized Proca (GP)
models, a vector tensor theory, in order to address the $H_{0}$
tension~\cite{Heisenberg:2014rta,DeFelice:2016yws}.

The GP theory is a ghost free vector tensor theory, with 5 degrees of
freedom, 3 from the massive gauge field sector which breaks
$U\left(1\right)$ symmetry and 2 from the gravity sector. In fact,
this theory propagates 1 scalar mode, 2 vector modes and 2 tensor
modes. This theory has equations of motion which are at most of second
order, in general curved space times~\cite{Heisenberg:2014rta}. The
cosmology of this theory which is quintic order in the Lagrangian
coupled with matter fluid was studied in~\cite{DeFelice:2016yws}. The
condition for the removal of both ghost instability and Laplacian
instability was found, in the high $k$ limit,
in~\cite{DeFelice:2016yws}. In this theory there exists a stable de
Sitter attractor with a dark energy equation of state
$w_{\text{DE}}=-1-s$ (during dust domination), where $s$ is a free
parameter in theory for the background. When $s=0$, the theory reduces
to $\Lambda\text{CDM}$ in the background level, but, as for
perturbation fields, this limit corresponds to strong coupling.

From the gravitational wave event
$\text{GW}170817$~\cite{TheLIGOScientific:2017qsa}, on combining it with
the gamma-ray burst $\text{GRB}170817\text{A}$
\cite{Goldstein:2017mmi}, the speed of propagation of the gravitational
waves $c_{T}$ is tightly constrained to be very close to $c$, where $c$ is the speed
of light. This restricts the GP Lagrangian to be, at most, of cubic
order. From the ISW cross correlation, the free parameter $s$ has the
best fit $s=0.185_{-0.089}^{+0.100}$~\cite{Nakamura:2018oyy}.  From
the previous study of observational constraints from the CMB shift
parameter, Baryon Acoustic Oscillation (BAO) and late time data,
i.e.\ Supernova, the background parameter $s$ is constraint to
$s=0.16\pm0.08$.  It was shown that the value of $H_{0}$ is compatible
with both early universe and late time cosmological data
sets~\cite{deFelice:2017paw}.  On the other hand, these previous works
were missing a few important points which are addressed here. In
particular, the constraints from Planck were coming only from a subset
of the Planck-data themselves, as previous works were only considering
the constraints on the CMB shift parameters. Although any viable model
needs to give a good fit to such observables, still, Planck data
consist of many other points, so that satisfying CMB-shift parameters
represents a necessary condition but not in general sufficient in
order for a given model to give a good fit to the Planck data.

Instead, in order to address this issue, in this work we make a full
analysis of the GP theory implementing both background and
perturbation in the Boltzmann code, CLASS~\cite{Blas:2011rf}, with
covariantly implemented baryon equations of motion
\cite{Pookkillath:2019nkn}.  For the background equations of motion we
make a backward integration with high enough precision for any
redshift needed to fit Planck data.  Then we perform an MCMC analysis
using Monte Python~\cite{Brinckmann:2018cvx,Audren:2012wb} (together
with Cosmomc~\cite{Lewis:2002ah}) against various cosmological data
sets, like Planck 2018, Hubble space telescope (HST), BAO, and Joint
Light-curve Analysis (JLA). We find that this theory reduces the
tension in the value of Hubble expansion rate 
$H_{0}=73.48_{-2.66}^{+2.56}$. This is in agreement with the previous
studies. On top of that, we find an extremely good fit with the data
sets, for example Planck 2018 + HST gives $\Delta\chi^{2}=22$, in
comparison with standard cosmology $\Lambda\text{CDM}.$ On the other
hand for data sets, JLA + Plank2018 + HST + BAO we get
$\Delta\chi^{2}=7$ of improvement with respect to
$\Lambda\text{CDM}$. This indicates that the GP theory reduces the
tension in $H_{0}$, and we will see later, the 2$\sigma$ regions for
GP and for the HST experiment do overlap. We also notice that the
value of $H_0$ reduces to $H_0= 70.41^{+0.94}_{-0.87}$ when including
BAO and JLA, which is still higher than the value obtained from the
same experiments but implementing the $\Lambda\text{CDM}$
model. Hence, as long as we consider both late time and early time
experiments, we can conclude that GP theory does not solve the $H_{0}$
tension, but it reduces it.

This paper is organized as follows. In section \ref{theory} we discuss
the GP theory and its background dynamics. In section \ref{pert}, we
discuss linear perturbation theory and determine the coupled equations
of motion for the perturbation field which can directly be implemented
in the Boltzmann code. Subsequently, we present our results in section
\ref{result}. We conclude our study in section \ref{conclu}.

\section{Theory\label{theory}}

The GP theory action is introduced
in~\cite{Heisenberg:2014rta,Heisenberg:2017mzp} and its cosmology is
studied in~\cite{DeFelice:2016yws}. With the constraints on the speed
of propagation of gravitational waves, i.e.\ $c_{T}=1$, the GP action
is given by,
\begin{equation}
S=\int
d^{4}x\sqrt{-g}\,(\mathcal{L}_{1}+\mathcal{L}_{2}+\mathcal{L}_{3}+\mathcal{L}_{{\rm
    m}})\,,
\end{equation}
where
\begin{eqnarray}
\mathcal{L}_{1} & = & \frac{\Mpl^{2}}{2}\,R\,,\\ \mathcal{L}_{2} & = &
-\frac{1}{4}\,F_{\mu\nu}\,F^{\mu\nu}+g_{2}(X)\,,\\ \mathcal{L}_{3} & =
& g_{3}(X)\nabla_{\mu}A^{\mu}\,,
\end{eqnarray}
where the field
$F_{\mu\nu}\equiv\partial_{\mu}A_{\nu}-\partial_{\nu}A_{\mu}$ and
$X\equiv-\frac{1}{2}\,A_{\mu}\,A^{\mu}$. For a concrete model of dark
energy, we assume the function $g_{2}\left(X\right)$ and
$g_{3}\left(X\right)$ of the form
\begin{equation}
g_{2}(X)=b_{2}\,X^{p_{2}}\,,\qquad
g_{3}(X)=b_{3}\,X^{p_{3}}\,,\label{eq:g2_g3}
\end{equation}
so that the background equations of motion have solution
\begin{eqnarray}
\varphi^{p}\,H & = & {\rm constant}\equiv\lambda\Mpl^{p}m,\label{eq:varphip_H}\\
  b_{2} & = & -m^{2}\Mpl^{2(1-p_{2})}\,,
\end{eqnarray}
where, since this theory in general breaks $U(1)$ gauge symmetry, we
set $A^{\mu}=(\varphi/a,0,0,0)$ on the background. Needless to say,
but the field $A^{\mu}$ does not represent the photon gauge vector
field. We have also introduced the Hubble factor as
$H\equiv\dot{a}/a^{2}$, where a dot represents here a derivative with
respect to the conformal time $\tau$. This theory was also discussed
in~\cite{deFelice:2017paw}.

The equation of motion for the field $\varphi$ has a solution if
\begin{eqnarray}
-\frac{1}{3} & = &
\frac{p_{3}b_{3}}{2^{p_{3}-p_{2}}p_{2}b_{2}}\,(\varphi^{p}H)\,,\\ p_{3}
& = & \frac{1}{2}\,(p+2p_{2}-1)\,.
\end{eqnarray}

For each matter Lagrangian $\mathcal{L}_{m}$ we have a perfect fluid
Lagrangian for which we have the energy-momentum tensor of the form
$T^{\mu}{}_{\nu}=\text{diag}\left(-\rho,P,P,P\right)$, which obeys the
conservation law
\begin{equation}
\dot{\rho}=-3H\left(\rho+P\right).
\end{equation}

\subsection{Background equations of motion}

In this section we briefly review the background equations of motion
considering an homogeneous and isotropic flat FLRW metric. The
calculation is following the same lines of~\cite{deFelice:2017paw}.

For general functions $g_{2}$, $g_{3}$ the background equations of
motion are given as,
\begin{align}
3\Mpl{}^{2}H^{2} &
=\rho_{A}+\sum_{i}\rho_{i}\,,\label{eq:FrEq1}\\ \Mpl^{2}\left(2\dot{H}+3H^{2}\right)
&
=-\sum_{i}P_{i}-P_{A}\,,\label{eq:FrEq2}\\ g_{3,X}+\frac{g_{2,X}}{3\varphi\,H}
& =0\,,\label{eq:BgEqPhi}
\end{align}
where
\begin{equation}
\rho_{A}=-g_{2},\qquad
P_{A}=-\varphi\left(t\right)^{2}\dot{\varphi}\left(t\right)g_{3}+g_{2}\,.
\end{equation}
The equation of state of the dark energy model is defined as
\begin{equation}
w_{\text{DE}}\equiv\frac{\rho_{A}}{P_{A}}=-1+\frac{\varphi\left(t\right)^{2}\dot{\varphi}\left(t\right)g_{3}}{g_{2}}.
\end{equation}
where $g_{2}$ and $g_{3}$ are defined by Eq.~\eqref{eq:g2_g3}, for the
concrete model of dark energy we are assuming. Notice that the
equation of state for dark energy deviates from the standard model of
$\Lambda\text{CDM}$. Now we need to parameterize this deviation from
$\Lambda\text{CDM}$.

Let us introduce $s=p_{2}/p$ and
\begin{equation}
\Omega_{{\rm DE}}\equiv\frac{1}{3\lambda^{2}2^{p_{2}}}\left(\frac{\varphi}{\Mpl}\right)^{2p(1+s)}\,,\label{eq:OmDE_def}
\end{equation}
then one can verify that
\begin{equation}
\Omega_{{\rm DE}}+\sum_{i}\frac{\rho_{i}}{3\Mpl^{2}H^{2}}=1\,.
\end{equation}
Also, let us make a convenient field redefinition 
CLASS code:
\begin{eqnarray}
\rho_{i} & = & 3\Mpl^{2}\varrho_{i}\,,\label{eq:rhoCL}\\ P_{i} & = &
3\Mpl^{2}p_{i}\,.\label{eq:pCL}
\end{eqnarray}
so that
\begin{equation}
H^{2}=\varrho_{A}+\sum_{i}\varrho_{i}\,,
\end{equation}
where
\begin{eqnarray}
\varrho_{A} & = &
\frac{1}{3\,2^{p_{2}}}\frac{H_{0}^{2}}{\lambda^{2}}\left[\frac{\varphi_{0}^{2p}}{\Mpl^{2p}}\right]\left(3\lambda^{2}2^{p_{2}}\Omega_{{\rm
    DE}}\right)^{\frac{s}{1+s}}\\ & = & H_{0}^{2}\left(\Omega_{{\rm
    DE}0}\right)^{\frac{1}{1+s}}\left(\Omega_{{\rm
    DE}}\right)^{\frac{s}{1+s}}\,.
\end{eqnarray}
To reach the first line of the above expression we used
Eq.~\eqref{eq:varphip_H}, Eq.~\eqref{eq:OmDE_def}. To get the final
expression we used Eq.~\eqref{eq:OmDE_def} to define
$\Omega_{\text{DE}0}$. When $s\to0$, $\rho_{A}$ becomes a constant,
this implies a $\Lambda$CDM limit for the background. From the
Friedmann equation one can see that
\begin{equation}
1=\Omega_{{\rm DE},0}+\sum_{i}\Omega_{i0}\,,
\end{equation}
so that $\Omega_{{\rm DE},0}$ is not a new parameter, but it can be
written in terms of the others.
$\Omega_{{\rm DE},0}\equiv\Omega_{\Lambda0}.$

From Eq.~\eqref{eq:varphip_H} and Eq.~\eqref{eq:OmDE_def}, we have
$H\propto\varphi^{-p}\propto\Omega_{{\rm DE}}^{-1/[2(1+s)]}$, or
\begin{eqnarray}
H & = & H_{0}\left(\frac{\Omega_{{\rm DE},0}}{\Omega_{{\rm
      DE}}}\right)^{1/[2(1+s)]},\\ \varphi & = &
\varphi_{0}\left(\frac{\Omega_{{\rm DE}}}{\Omega_{{\rm
      DE},0}}\right)^{1/[2p(1+s)]},
\end{eqnarray}
and
\begin{equation}
\frac{\varphi_{0}}{\Mpl}=\left(\frac{\lambda m}{H_{0}}\right)^{1/p}.
\end{equation}
On evaluating $\varrho_{A}(a=1)$, we also get
\begin{equation}
\frac{m^{2}}{3\,2^{p_{2}}}\left(\frac{\varphi_{0}}{\Mpl}\right)^{2ps}=H_{0}^{2}\,\Omega_{{\rm
    DE}0}\,,
\end{equation}
or
\begin{equation}
\frac{m}{H_{0}}=\sqrt{3\,2^{p_{2}}\Omega_{{\rm
      DE},0}}\left(\frac{\Mpl}{\varphi_{0}}\right)^{ps},
\end{equation}
so that
\begin{equation}
\frac{\varphi_{0}}{\Mpl}=\left(\lambda\sqrt{3\,2^{p_{2}}\Omega_{{\rm
      DE},0}}\right)^{1/[p(1+s)]}.
\end{equation}
This relation can be used to define $\lambda$ and $m$ in terms of
$\varphi_{0}$ and the other variables.

Along the same lines, one can see that the second Einstein equation
can be written as
\begin{equation}
\frac{2}{3}\,\frac{\dot{H}}{a}+H^{2}+\sum_{i}p_{i}+p_{A}=0\,,
\end{equation}
which, once we replace $H=H(\varphi$) can be solved for
$\dot{\varphi}$ in terms of the other variables. In this case we find
that, in terms of the time-independent variable $N=\ln(a),$ the
background equations of motion can be written as
\begin{eqnarray}
\Omega'_{{\rm DE}} & = & \frac{(1+s)\,\Omega_{{\rm
      DE}}\,(3+\Omega_{r}-3\Omega_{{\rm DE}})}{1+s\Omega_{{\rm
      DE}}}\,,\\ \Omega'_{r} & = &
-\frac{\Omega_{r}\,[1-\Omega_{r}+(3+4s)\Omega_{{\rm
        DE}}]}{1+s\Omega_{{\rm DE}}}\,.
\end{eqnarray}
As a result we can solve for any given value of $a$ (or $N$) the value
of $\Omega_{D}E$ and $\Omega_{r}$, whereas $\Omega_{m}=1-\Omega_{{\rm
    DE}}-\Omega_{r}$.

\section{Perturbations\label{pert}}

Now, let us look at the behaviour of the perturbation fields of the
theory around a flat FLRW metric. Linear perturbation around flat FLRW
metric and ghost condition is studied in ~\cite{deFelice:2017paw}.
However, to implement the equations of motion in Boltzmann code, we
have to express the equations of motion in a fashion, which is
suitable for the Boltzmann code being used (in this we use
CLASS~\cite{Blas:2011rf}).  This is explained briefly in the
following.

We adopt the usual technique for finding the linear perturbation
equations of motion by expanding action up to second order in
perturbation variables, without choosing any gauge. Only after finding
the equations of motion for each of the fields we choose a gauge. Then
we construct linear combinations of the previously obtained equations
of motion, and perform convenient field redefinitions in order to find
suitable equations of motion which can be easily implemented in the
Boltzmann code. Since the expressions are quite long, we only outline
our calculations below. We report the final expressions of the new
equations of motion to be implemented in the code in Appendix
\ref{pert_eoms}.

In the following we consider the flat FLRW metric with perturbations
\begin{equation}
ds^{2}=-a^{2}\,\left(1+2\alpha\right)\,d\tau^{2}+2a\,\partial_{i}\chi\,d\tau\,dx^{i}+a^{2}\,\left[\left(1+2\zeta\right)\,\delta_{ij}+\partial_{i}\partial_{j}E/a^{2}\right]\,dx^{i}\,dx^{j}\,,\label{eq:FLRW}
\end{equation}
and we introduce matter fields in the usual way, because each matter
field has no coupling with the Proca field. As such, we use the matter
Lagrangian of the form as discussed
in~\cite{Pookkillath:2019nkn,Schutz:1977df,DeFelice:2009bx}
\begin{equation}
S_{m}=-\int
d^{4}x\,\sqrt{-g}[\rho(n,s)+J_{m}^{\mu}\,(\partial_{\mu}l)]\,,
\end{equation}
where $\rho$ is matter energy density, $n$ number density of the
matter species. The other fundamental variables are the timelike
vector $J_{m}^{\alpha}$, the metric $g_{\mu\nu}$, and the scalar $l$,
whereas:
\begin{equation}
n\equiv\sqrt{-J_{m}^{\mu}J_{m}^{\nu}g_{\mu\nu}}\,.\label{eq:def_n}
\end{equation}
At linear order in perturbation theory about an FLRW background
Eq.~\eqref{eq:FLRW}, one can define as follows:
\begin{align}
l & =-\int_{0}^{\tau}d\eta\,a(\eta)\,\overline{\rho}_{,n}+\delta
l\,,\\ J_{m}^{0} & =\frac{N_{0}}{a^{4}}(1+W_{0})\,,\\ J_{m}^{i} &
=\frac{\partial_{j}W}{a^{2}}\,\delta^{ij}\,,
\end{align}
where $\bar{\rho}_{,n}\equiv\frac{\partial\bar{\rho}}{\partial n}$.
We have also the vector field $A^{\mu},$whose components will be
written as
\begin{eqnarray}
A^{0} & = & \frac{\varphi(t)+\delta\varphi}{a}\,,\\ A^{i} & = &
\frac{1}{a^{2}}\,\delta^{ik}\,\partial_{k}J\,,
\end{eqnarray}
where we consider here only scalar perturbations. As shown
in~\cite{DeFelice:2016yws}, the vector modes do not affect the
evolution of the matter fields vector modes which still show the usual
decaying behaviour. Notice that so far, we have not set yet any
gauge. After expanding the action at second order in the
perturbations, we can find equations of motion for each of the
perturbation field. From the gravity sector we have 4 equations of
motion, 2 for the vector modes and the remaining equations of motion
for each matter component. We also redefine the matter field variables
as
\begin{align}
\delta l & =\rho_{,n}v\\ W_{0} &
=\frac{\rho}{n\rho_{,n}}\delta-\alpha\\ v &
=-\frac{a}{k^{2}}\,\theta\,,
\end{align}
for each of the matter component.

Once we have the equations of motions for every field we fix a gauge.
In the following we study the Newtonian gauge case, so that we set
\begin{eqnarray}
\alpha & = & \psi\,,\\ \chi & = & 0\,,\\ \zeta & = & -\phi\,,\\ E & =
& 0\,.
\end{eqnarray}

By combining equations of motion for $E$ and $\zeta$, we get the same
one for GR, which can be used to solve $\psi$ in terms of $\phi$ and
the shear $\sigma$, as in
\begin{eqnarray}
\psi & = &
\phi-\frac{9}{2}\,\frac{a^{2}}{k^{2}}\,\Gamma_{\sigma}\,,\label{eq:shearGR}\\ \Gamma_{\sigma}
& \equiv & \sum_{i}(\varrho_{i}+p_{i})\,\sigma_{i}\,.
\end{eqnarray}

Therefore this gravitational equation does not get any modification,
or, in other words, the GP theory does not affect the gravitational
shear.

Now we still need to find another equation of motion to fix $\phi$
together with the new degrees of freedom coming from the Proca action.
For this goal, we can use the EOMs for $\chi$, $\delta A$, and
$\delta\varphi$ in order to set a dynamics for the remaining
gravity/vector fields.  In order to make these EOMs first order ODEs,
it is useful to perform the following field redefinition
\begin{eqnarray}
J & = & J_{2}-\frac{\varphi}{pH}\,\phi\,,\\ \delta\varphi & = &
\delta\varphi_{2}-2\varphi\,\psi-\frac{1}{a}\,\frac{dJ}{d\tau}\,,
\end{eqnarray}
so that in this case the equation of motion for $\delta\varphi$,
$E_{\delta\varphi}$ only depends on $\delta\varphi_{2}$, and no time
derivative for fields, except for $\dot{J}_{2}$. Therefore we can
solve for $\delta\varphi$ in terms of the other variables.

We can now see that the equation for $J_{2}$, i.e.\ the equation
$E_{J}$, now becomes a second order ODE for $J_{2}$, and can be
written in terms of
\begin{equation}
E_{J}=E_{J}(\ddot{J}_{2},\dot{J}_{2},J_{2},\dot{\phi},\phi,\dot{\psi},\psi)=0\,,
\end{equation}
which can be rewritten as an ODE for $\ddot{J}_{2}$. In order to do
this we need also to replace the EOMs for $\dot{\phi}$ and
$\dot{\psi}$
\footnote{The contribution coming from the $\dot{\psi}$ term in this
  equation of motion is proportional to $\varphi$ as in
  $\ddot{J}_{2}\propto a\varphi\dot{\psi}$.  Therefore such a term is
  negligible at early times, since, in this case, $\Omega_{{\rm
      DE}}\to0$, and, as we shall we see later on,
  $\varphi^{2}/(H^{2}\Omega_{{\rm DE}})\to0$. Instead, at late times,
  when the Proca contributions play some non-trivial role, then photon
  shear becomes more and more negligible, so that we can consider
  $\dot{\psi}\approx\dot{\phi}$ as a sensible approximation.}. From a
linear combination of $E_{J}$ and $E_{\chi}$, we find a new equation of
motion, $E_{J\chi},$ which can be written as
\begin{equation}
E_{J\chi}=E_{J\chi}(\dot{\phi},\phi,\dot{J}_{2},J_{2},\psi,\sum_{i}(\rho_{i}+p_{i})\,\theta_{i})\,,
\end{equation}
which reduces to the standard momentum equation in the limit $s\to0$.
Therefore, we can now solve all the equations of motion for the
variables $\dot{p}_{2}$, $p_{2}=\dot{J}_{2}$, $\dot{\phi}$, and
$\psi$.

Finally let us consider the prior conditions we can get for the
parameters in the theory coming from the no-ghost condition. On
studying the propagation for the new Proca scalar mode, one finds the
no-ghost condition
\begin{equation}
Q=\frac{3\Mpl^{2}p^{2}sH^{2}\Omega_{{\rm DE}}\,(\Omega_{{\rm
      DE}}s+1)}{(\Omega_{{\rm DE}}ps-1)^{2}\,\varphi^{2}}\,.
\end{equation}
We notice that we cannot set $p^{2}s$ to vanish, otherwise the mode
would become strongly coupled. This implies that $s>0$, and since
$\Omega_{{\rm DE}}>0$, then we can see that $Q>0$, so that no ghost
exists during the evolution of the universe. However we need to make
sure that at early times we still avoid strong coupling, i.e.~$Q\to0$.
Since $\varphi^{2}\propto\Omega_{{\rm DE}}^{1/[p(1+s)]}$, and
$H^{2}\propto\Omega_{{\rm DE}}^{-1/(1+s)}$, we find that as
$\Omega_{{\rm DE}}\to0$ that
\begin{equation}
Q\sim\Omega_{{\rm DE}}^{1-1/(1+s)-1/[p(1+s)]}=\Omega_{{\rm
    DE}}^{(ps-1)/[p(1+s)]}\,,
\end{equation}
so that we require
\begin{equation}
ps-1<0\,,
\end{equation}
or $0<ps<1$, so that the field becomes at most weakly coupled at early
times. This condition implies, at early times, that
\begin{equation}
\frac{\varphi^{2}}{H^{2}\Omega_{{\rm DE}}}\sim\Omega_{{\rm
    DE}}^{(1-ps)/[p(1+s)]}\to0\,.
\end{equation}
In the code, we will define $\bar{J}=J_{2}/\Mpl$, and
$\bar{\varphi}=\varphi/\Mpl$.

As initial conditions, at very large redshifts, since the Proca
contributions become more and more negligible, it is sensible to
consider the following initial conditions $J=0=\delta\varphi$. In this
case we also find
\begin{equation}
\bar{J}_{{\rm ini}}=\frac{J_{2,{\rm
      ini}}}{\Mpl}=\frac{\varphi/\Mpl}{pH}\,\phi_{{\rm
    ini}}\propto\Omega_{{\rm DE}}^{(1+p)/[2p(1+s)]}\to0\,,
\end{equation}
so that we will also consider the case $\bar{J}=0=\dot{\bar{J}}.$


\section{Results\label{result}}

In this section we present our results after the MCMC analysis. We run
the Boltzmann code for the GP theory and find the cosmological
constraints to the parameters with Planck + HST, and also with Planck
+ HST + BAO + JLA. We fit the GP theory and $\Lambda$CDM from the
observational data with the MCMC method. The dataset includes those of
the CMB temperature fluctuation from \textit{Planck 2018} with
\texttt{Planck\_highl\_TTTEEE}, \texttt{Planck\_lowl\_EE},
\texttt{Planck\_lowl\_TT}, \texttt{Planck\_lensing}
polarization~\cite{Aghanim:2019ame,Akrami:2019izv,Aghanim:2018eyx,Akrami:2018odb,Aghanim:2018oex},
the single data point of the Hubble constant
$H_{0}=74.03_{-1.42}^{+1.42}$ from Hubble Space Telescope (HST)
observations~\cite{Riess:2019cxk}, the baryon acoustic oscillation
(BAO) data from 6dF Galaxy Survey~\cite{Beutler:2011hx} and the Sloan
Digital Sky Survey~\cite{Ross:2014qpa,Alam:2016hwk}.  The joint light
curves (JLA) comprised of 740 type Ia supernovae from
\cite{Betoule:2014frx}. We find that, as for the GP background, the
$H_{0}$ tension is completely removed between early universe and late
time measurements, when Planck + HST data alone are considered. The
value estimated for $H_0$ reduces on introducing the intermediate data
set BAO and JLA, but still higher that what is estimated from
$\Lambda\text{CDM}$. Hence, reducing the tension for this
measurement. The constraints for the parameters of the GP theory as
well as for the $\Omega_{m}$ and $H_{0}$ at $95\%$ C.L.\ are given. We
also notice that for this theory there is a good improvement in the
$\chi^{2}$ value in comparison with $\Lambda\text{CDM}$ model of
cosmology. The priors of the various cosmological parameters are
listed in Table~\ref{tab:prior}. Here, we have set $s$ to be a
positive number as to avoid ghost (and strong coupling) in the scalar
mode.

\begin{table}[ht]
\centering{}\caption{ Priors for cosmological parameters in
  Generalized Proca Theory. }
\begin{tabular}{|c||c|}
\hline \textbf{Parameter} & \textbf{Prior} \tabularnewline \hline $s$
& $0\leq s\leq0.99$ \tabularnewline \hline $p_{2}$ & $10^{-2}\leq
p_{2}\leq0.99$ \tabularnewline \hline $\log_{10}{\bar{\varphi}_{0}}$ &
$-2.5\leq\log_{10}{\bar{\varphi}_{0}}\leq2$ \tabularnewline \hline
\end{tabular}
 \label{tab:prior} 
\end{table}

\subsection{Planck + HST}

\begin{figure}[ht]
\includegraphics[width=9cm]{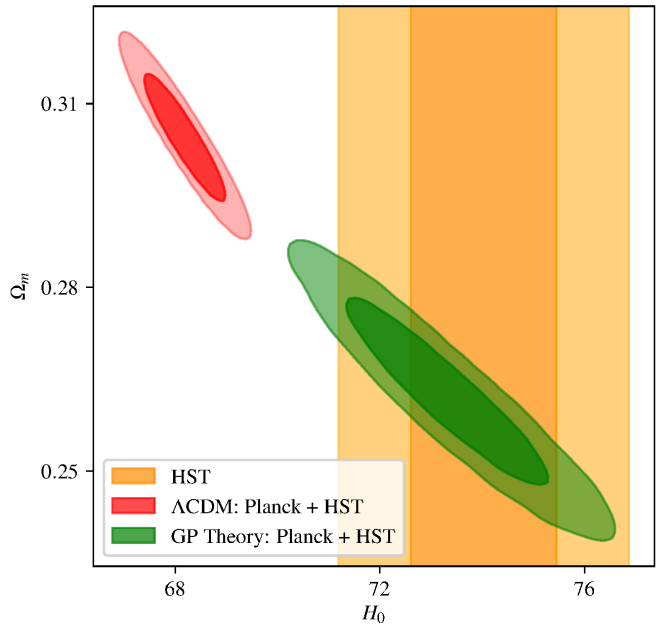} \caption{Combining
  all the data together, the GP model is able to make $H_{0}$
  measurements compatible at 2-sigma with the MCMC results.}
\label{fig:H0-2sigma} 
\end{figure}

In this subsection we show cosmological constraints of the parameters
with Planck + HST. The value of $H_{0}$ that is derived from the GP
theory gives a higher value in comparison with that of
$\Lambda\text{CDM}$.  This value perfectly matches with that of local
distance ladder measurement with $H_{0}=$$73.34_{-2.69}^{+2.46}$ at
$95\%$ C.L. Hence the tension in the value of $H_{0}$ is removed
within the GP theory as shown in Fig.\ \ref{fig:H0-2sigma}.

\begin{figure}
\centering \includegraphics[width=0.8\linewidth]{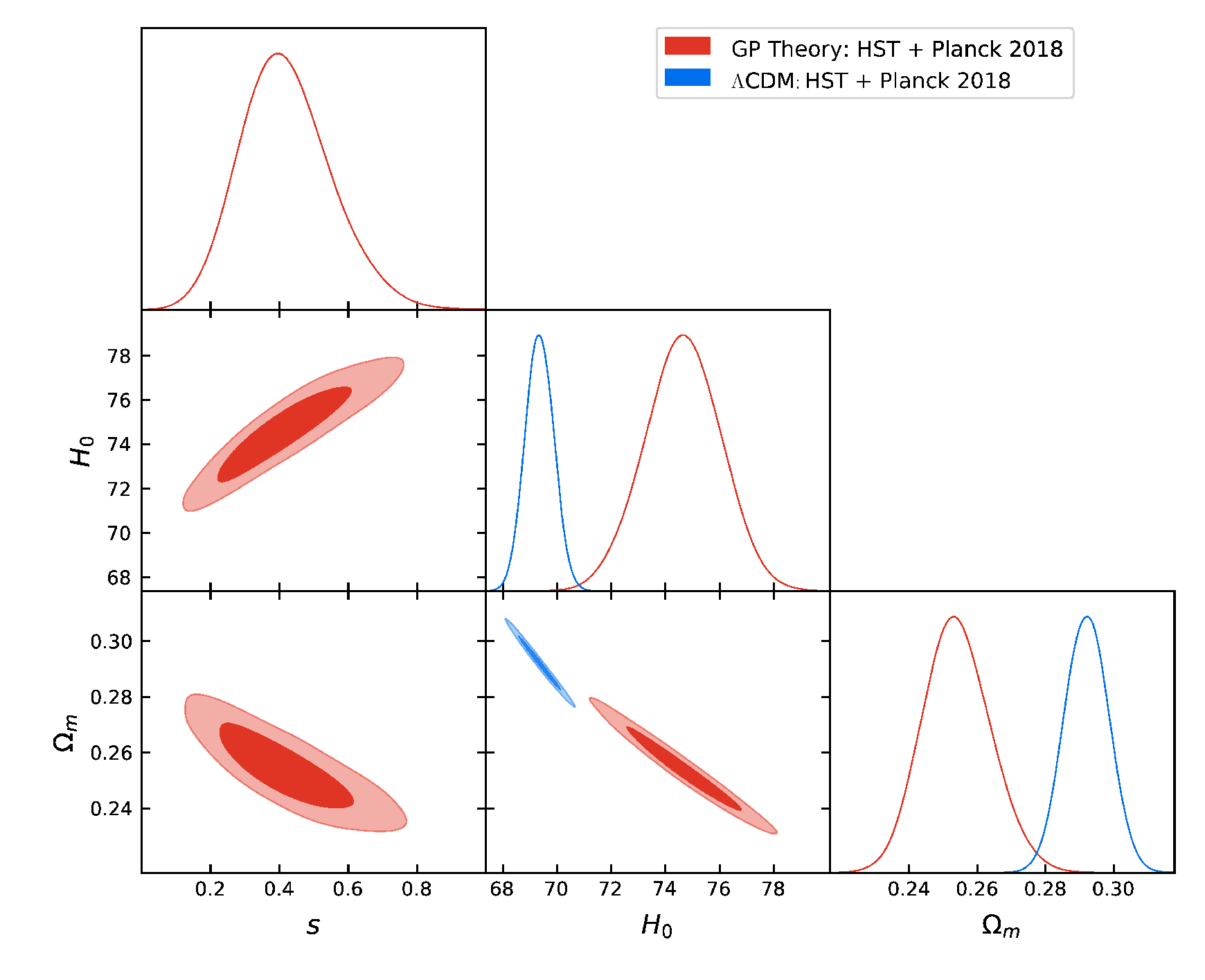} \caption{One
  and two-dimensional distributions of $s$, $H_{0}$ and $\Omega_{m}$
  in Planck+ HST data sets, where the contour lines represent 68$\%$~
  and 95$\%$~ C.L., respectively. The results obtained here via
  CosmoMC are shown to be consistent with the ones obtained by
  CLASS/Montepython.  This test gives a strong check on the
  consistency of (either of) the code.}
\label{fg:background-planck+hst} 
\end{figure}

We also show in Fig.\ \ref{fg:background-planck+hst} the results for
the same data sets having followed a slightly different approach by
using CosmoMC. We can see that the two results are completely
consistent.  This result gives a check for the consistence of (either
of) the code.

\begin{figure}[ht]
\includegraphics[width=15cm]{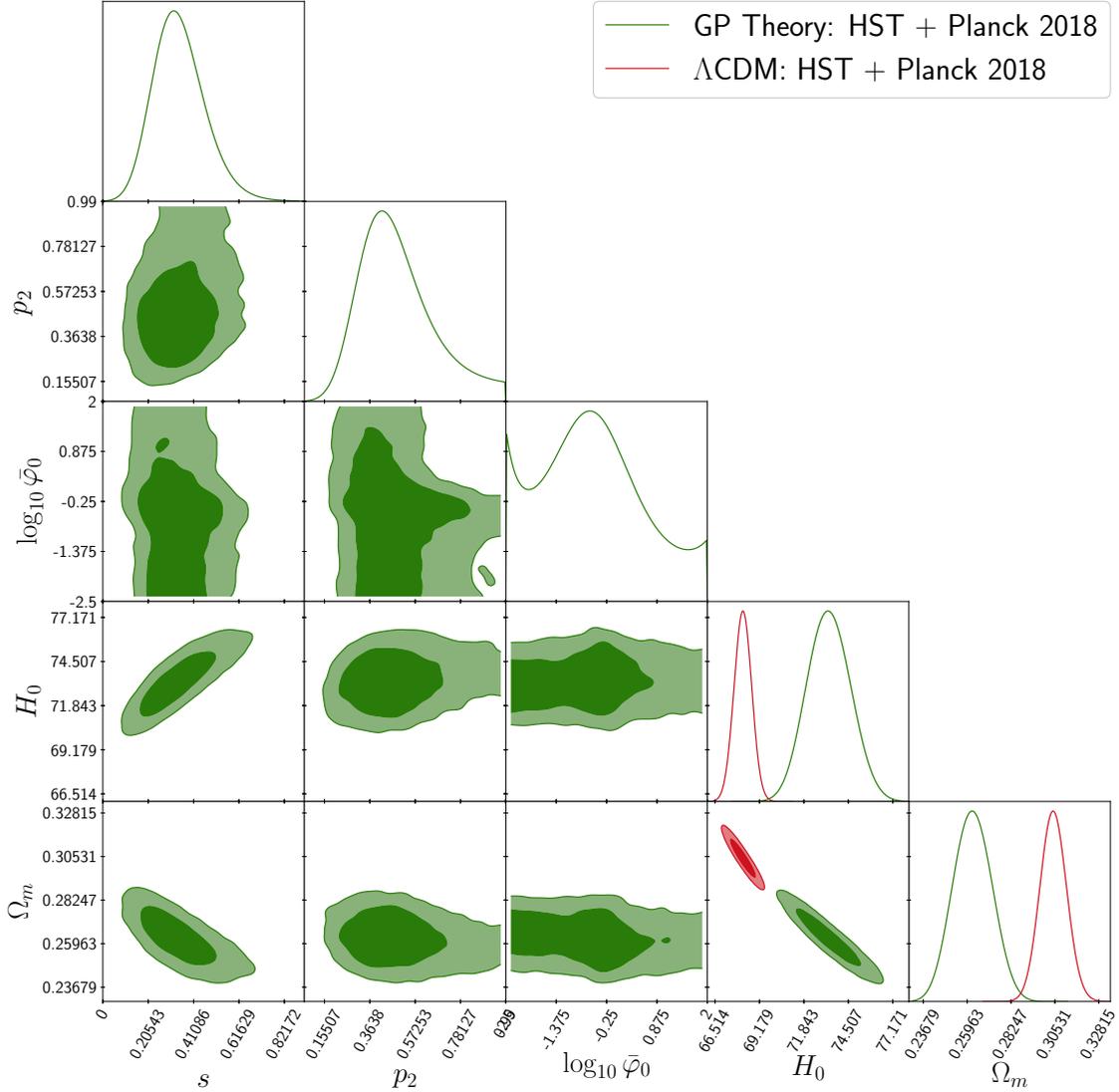}
\caption{The triangular plot for the cosmological parameters for the
  Generalized Proca theory confronted with Planck + HST data sets.}
\label{fig:pla+hst_tri} 
\end{figure}

Table~\ref{tab:pla+hst} shows the constraints to the parameters up to
$95\%$ of C.L. Notice that for the parameter $\bar{\varphi}_{0}$ we
could only give an upper bound (at 1-sigma). This is also shown in the
triangular plot Fig.\ \ref{fig:pla+hst_tri}.

\begin{table}[ht]
\begin{tabular}{|l|c|c|c|c||c|c|c|c|}
\hline & \multicolumn{4}{c||}{\textbf{Generalised Proca}} &
\multicolumn{4}{c|}{\textbf{$\boldsymbol{\Lambda\text{CDM}}$}}\tabularnewline
\hline Param & best-fit & mean$\pm\sigma$ & 95\% lower & 95\% upper &
best-fit & mean$\pm\sigma$ & 95\% lower & 95\% upper\tabularnewline
\hline $s$ & $0.3342$ & $0.3455_{-0.14}^{+0.095}$ & $0.1233$ & $0.592$
& - & - & - & -\tabularnewline \hline $p_{2}$ & $0.8344$ &
$0.4921_{-0.21}^{+0.11}$ & $0.1885$ & $0.8783$ & - & - & - &
-\tabularnewline \hline $\log_{10}{\bar{\varphi}_{0}}$ & $-0.7579$ &
$-0.5604_{nan}^{+nan}$ & $nan$ & $nan$ & - & - & - & -\tabularnewline
\hline $H_{0}$ & $73.48$ & $73.34_{-1.3}^{+1.3}$ & $70.79$ & $75.94$ &
$68.26$ & $68.18_{-0.53}^{+0.53}$ & $67.13$ & $69.25$\tabularnewline
\hline $\Omega_{m}$ & $0.2607$ & $0.2625_{-0.01}^{+0.0099}$ & $0.2429$
& $0.2824$ & $0.3039$ & $0.3045_{-0.0071}^{+0.0069}$ & $0.2907$ &
$0.3186$\tabularnewline \hline
\end{tabular}

\caption{Cosmological constraints for the parameters at 1$\sigma$ and 2$\sigma$ for both Generalized Proca
  theory and $\Lambda\text{CDM}$ confronted with Planck+HST data sets.}
\label{tab:pla+hst} 
\end{table}

For the GP theory, the bestfit value gives $\chi^{2}=2773$, whereas
$\Lambda\text{CDM}$ gives
$\chi^{2}=2794.64$. Table~\ref{hst_planck_chi2eff} shows the effective
$\chi^{2}$ for the individual experiments. That is, there is a
remarkable betterment in the fitting of the GP theory in comparison
with that of $\Lambda\text{CDM}$ with $|\Delta\chi^{2}|\sim22$.  We
also give the triangular plots of the parameters in
Fig.~\ref{fig:pla+hst_tri}.

\begin{table}[H]
\begin{centering}
\begin{tabular}{|c||c|c|}
\hline \multirow{2}{*}{\textbf{Experiments}} &
\multicolumn{2}{c|}{\textbf{$\chi^{2}$ effective}}\tabularnewline
\cline{2-3} \cline{3-3} & \textbf{Generalised Proca } &
\textbf{$\boldsymbol{\Lambda\text{CDM}}$}\tabularnewline \hline \hline
Planck\_highl\_TTTEEE & 2346.51 & 2348.98\tabularnewline \hline
Planck\_lowl\_EE & 395.67 & 397.03\tabularnewline \hline
Planck\_lowl\_TT & 21.82 & 22.87\tabularnewline \hline Planck\_lensing
& 8.45 & 9.21\tabularnewline \hline hst & 0.15 & 16.54\tabularnewline
\hline Total & 2772.60 & 2794.64\tabularnewline \hline
\end{tabular}
\par\end{centering}
\caption{$\chi^{2}$ effective for individual experiments with the
  dataset of Planck + HST.}
\label{hst_planck_chi2eff} 
\end{table}

\subsection{Planck, JLA, HST and BAO}

\begin{figure}[ht]
\includegraphics[width=13cm]{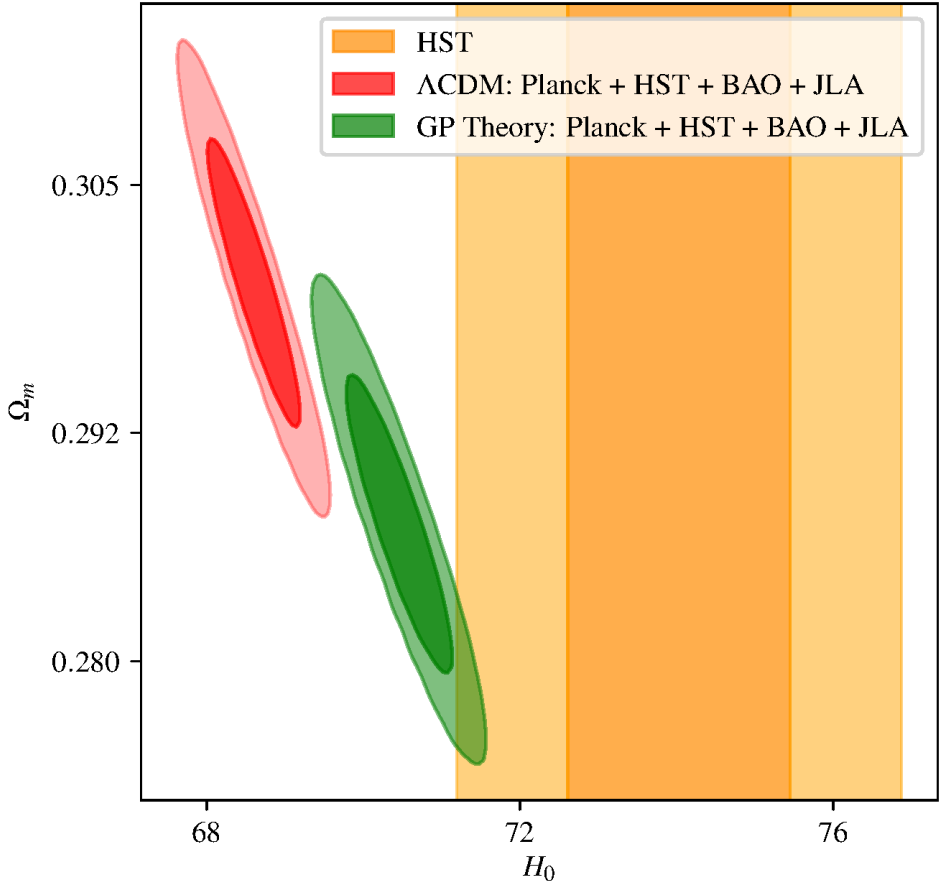} \caption{Combining
  all the data together, the GP model is able to make $H_{0}$
  measurements compatible at 2-sigma with the MCMC results.}
\label{fig:H0-2sigma-all} 
\end{figure}

We also check the cosmological constraints for the GP theory combining
Planck, JLA, HST and BAO. Table~\ref{tab:all_data} shows that
constraints to the parameters and a comparison with the
$\Lambda\text{CDM}$ values for $H_{0}$ and $\Omega_{m}$. We have found
that the value of $s$ on using all the data is reduced. It is
interesting to notice that the value of the $H_{0}$ has changed with
in the $95\%$ C.L.\ in comparison with $\Lambda\text{CDM}$. However,
within the GP theory, Planck data and local measurement of $H_{0}$
agree within 2 sigma (by this we mean that the 2$\sigma$ regions for
Proca and the $H_{0}$ measurements overlap), as shown in
Fig.\ \ref{fig:H0-2sigma-all}.

Therefore, GP is able to reduce the tension in the data we have
considered.  This behaviour sounds really promising and it should be
checked against future data. 

The parameter $\bar{\varphi}_{0}$ as in the previous case has a large
degeneracy, and we can only give an upper bound (at 1-sigma) to it.

\begin{table}[ht]
\begin{tabular}{|l|c|c|c|c||c|c|c|c|}
\hline & \multicolumn{4}{c||}{\textbf{Generalised Proca}} &
\multicolumn{4}{c|}{\textbf{$\boldsymbol{\Lambda\text{CDM}}$}}\tabularnewline
\hline Param & best-fit & mean$\pm\sigma$ & 95\% lower & 95\% upper &
best-fit & mean$\pm\sigma$ & 95\% lower & 95\% upper\tabularnewline
\hline $s$ & $0.172$ & $0.1722_{-0.01}^{+0.013}$ & $0.1432$ & $0.1984$
& - & - & - & -\tabularnewline $p_{2}$ & $0.226$ &
$0.232_{-0.041}^{+0.038}$ & $0.1392$ & $0.3246$ & - & - & - &
-\tabularnewline $\log_{10}{\bar{\varphi}_{0}}$ & $-0.2856$ &
$-1.492_{nan}^{+nan}$ & $nan$ & $nan$ & - & - & - & -\tabularnewline
$H_{0}$ & $70.41$ & $70.45_{-0.45}^{+0.45}$ & $69.54$ & $71.35$ &
$68.59$ & $68.6_{-0.4}^{+0.4}$ & $67.79$ & $69.4$\tabularnewline
$\Omega_{m}$ & $0.2877$ & $0.2873_{-0.0054}^{+0.0051}$ & $0.2768$ &
$0.2977$ & $0.3001$ & $0.2999_{-0.0052}^{+0.005}$ & $0.2896$ &
$0.3102$\tabularnewline \hline
\end{tabular}

\caption{This table shows the cosmological constraints to the
  parameters for both the generalized Proca theory and
  $\Lambda\text{CDM}$ confronted with Planck + HST + BAO + JLA data
  sets.}
\label{tab:all_data} 
\end{table}

\begin{figure}[ht]
\includegraphics[width=15cm]{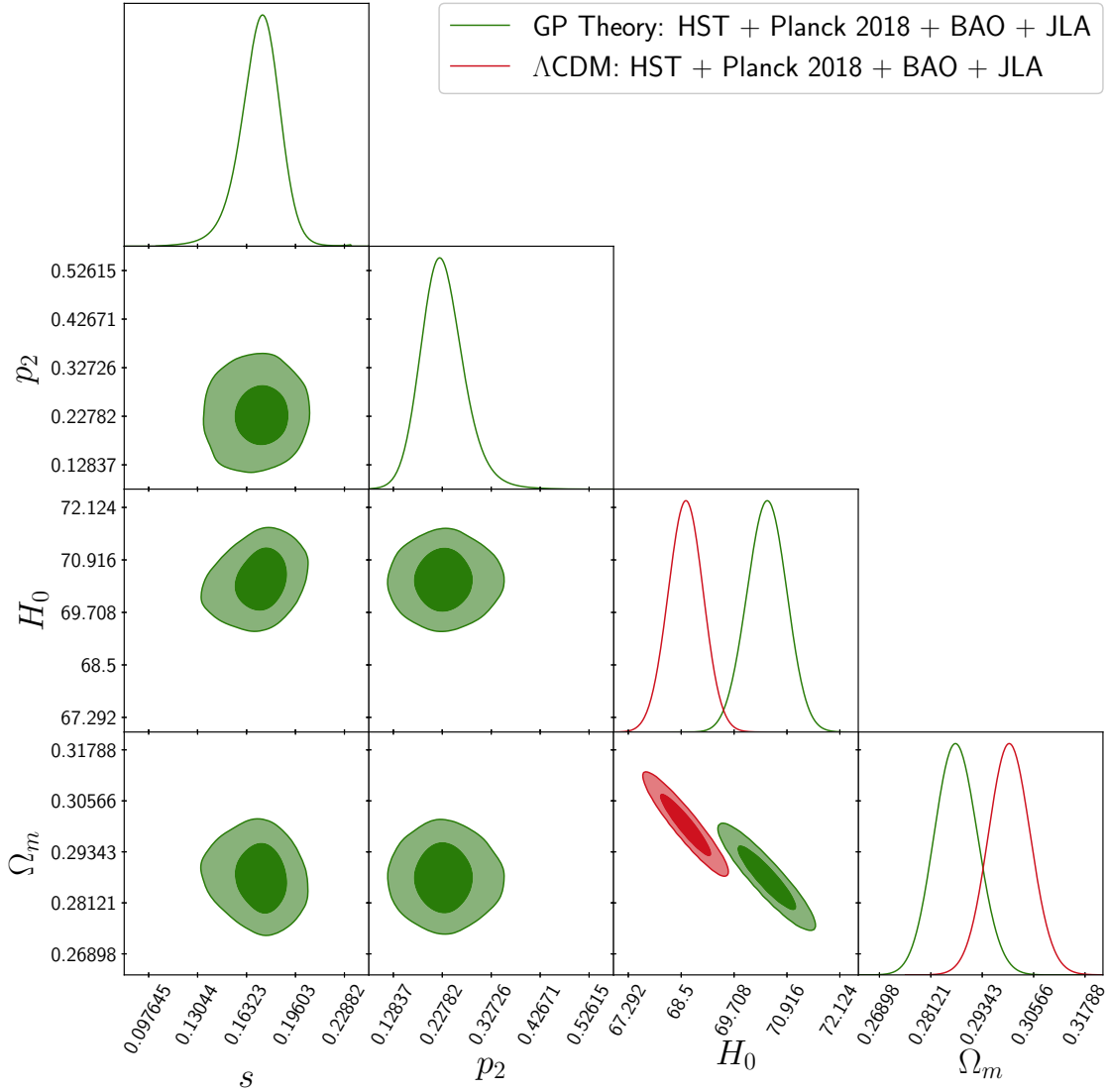}

\caption{The triangular plot for the cosmological parameters for the
  Generalized Proca theory confronted with Planck + HST + BAO + JLA
  data sets.}
\label{fig:all_tria} 
\end{figure}

For this theory we get a $\chi^{2}=3472$, which is again lower than
that of $\Lambda\text{CDM}$, which is $\chi^{2}=3479$, resulting in
difference of $|\Delta\chi^{2}|=7$. Table~\ref{all_data_chi2eff} shows
the effective $\chi^{2}$ for the individual experiments. For this data
set also we find that there is a preference of the GP theory over
$\Lambda\text{CDM}$. It is evident from the
table~\ref{all_data_chi2eff}, that the improvement in effective
$\chi^2$ of HST is done at the cost of a (partial) degradation of the
effective $\chi^2$ of BAO and JLA. We show the triangular plot for the
parameters in Fig.\ \ref{fig:all_tria}. Besides noticing the
difference of $\chi^2$, we want to consider other statistical analysis
to compare the two models. In particular, we select not only Akaike
Information Criterion (AIC) but also AIC with a correction (which
under certain conditions, is more accurate, especially for small
sample sizes) (AICc) and Bayesian Information Criterion (BIC) to
compare the two models~\cite{Trotta:2008qt}. The AICc criterion can be
expressed by
 \begin{equation}
 AICc \equiv -2 \ln L_{max}+2k + \frac{2k^2+2k}{n-k-1},
 \end{equation}
 where $L_{max}\equiv p(d|\theta_{max}, M)$ is the maximum likelihood
 value and $k$ is the number of free parameters in the model. Here,
 $n$ denotes the sample size of the simulation and it will lead AICc
 to converge to AIC when $n\to\infty$. Since $\chi^2=-2 \ln L_{max}$
 and GP just has one more free background parameter, $s$, than
 $\Lambda$CDM, we can get
 $\triangle AICc= \triangle AICc_{\Lambda CDM}-\triangle AICc_{GP}=
 20.031 (4.321)$ in Planck+HST (Planck+HST+BAO+JLA) dataset. Since there
 is a large degeneracy for $\bar{\varphi}_{0}$ we can, without loss of
 generality, fix it to some value. On adding the perturbation parameter
 $p_2$, we find that
 $\triangle AICc= \triangle AICc_{\Lambda CDM}-\triangle
 AICc_{GP}= 18.02 (2.31)$. Even in this case, the value for $\triangle AICc$ shows preference
 for the GP theory compared to the $\Lambda$CDM  model~\cite{Arevalo:2016epc}.

On the other hand, the Bayesian criterion is defined as
 \begin{equation}
BIC \equiv -2 \ln L_{max}+k \ln n.
\end{equation}
The $\triangle BIC= \triangle BIC_{\Lambda CDM}-\triangle BIC_{GP}=
5.65$. It shows the evidence against for our GP theory more close to
the observational data than the $\Lambda$CDM in Planck+HST
dataset. This result is consistent with our conclusion from the
difference of $\chi^2$ in global fitting.

\begin{table}[H]
\begin{centering}
\begin{tabular}{|c||c|c|}
\hline \multirow{2}{*}{\textbf{Experiments}} &
\multicolumn{2}{c|}{\textbf{$\chi^{2}$ effective}}\tabularnewline
\cline{2-3} \cline{3-3} & \textbf{Generalised Proca } &
\textbf{$\boldsymbol{\Lambda\text{CDM}}$}\tabularnewline \hline \hline
Planck\_highl\_TTTEEE & 2343.12 & 2346.06\tabularnewline \hline
Planck\_lowl\_EE & 395.72 & 397.64\tabularnewline \hline
Planck\_lowl\_TT & 22.43 & 23.01\tabularnewline \hline Planck\_lensing
& 9.59 & 8.74\tabularnewline \hline JLA & 685.36 &
683.04\tabularnewline \hline bao\_boss\_dr12 & 5.60 &
3.39\tabularnewline \hline bao\_smallz\_2014 & 3.87 &
2.03\tabularnewline \hline hst & 6.51 & 14.65\tabularnewline \hline
Total & 3472.22 & 3478.55\tabularnewline \hline
\end{tabular}
\par\end{centering}
\caption{$\chi^{2}$ effective for individual experiments with the
  dataset of Planck + HST + BAO +JLA.}
\label{all_data_chi2eff} 
\end{table}

\begin{figure}[ht]
\includegraphics[width=14cm]{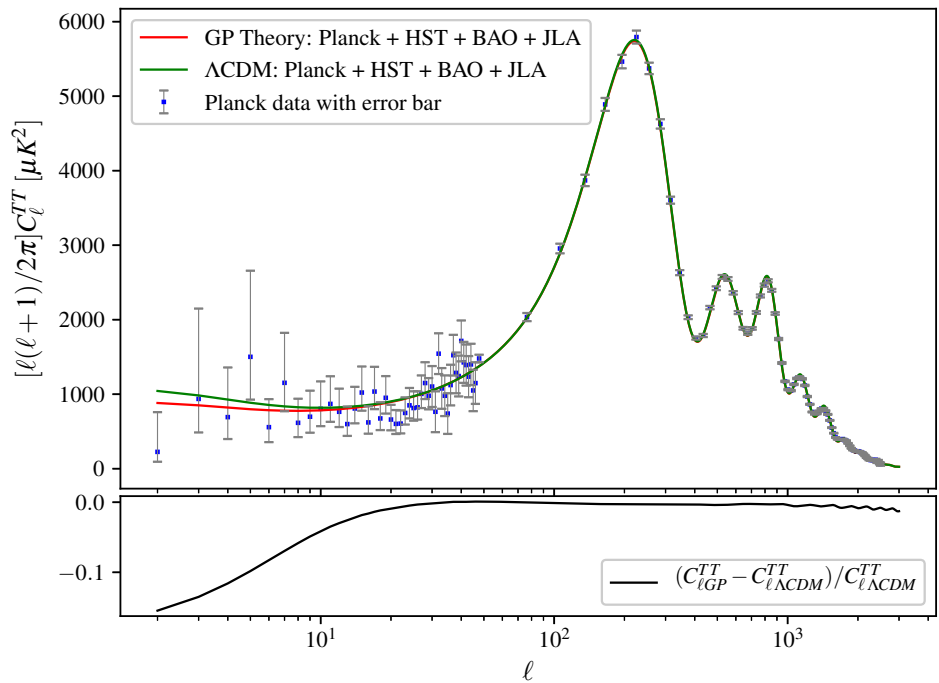}

\caption{CMB power spectra TT mode for the $\Lambda$CDM and GP Theory
  obtained using the bestfit parameters for both theories to the
  Planck + HST + BAO + JLA data set. The lower panel shows the
  relative difference of the $C_{\ell}^{TT}$ of GP theory with respect
  to $\Lambda$CDM. The error-bars are $1-\sigma$ from Planck 2018 TT
  power spectra.}
\label{fig:cmbTT} 
\end{figure}

In order to show the modification of GP Theory in the early universe,
we compare the difference of CMB power spectra of the TT mode for GP
Theory and $\Lambda$CDM with the observation we used above. In
Fig.~\ref{fig:cmbTT}, the $\Lambda$CDM and GP Theory with the value of
parameters are form best-fit value of Planck + HST + BAO + JLA data
set. In the high-$\ell$ part of Fig.~\ref{fig:cmbTT}, we can find the
GP is very close to $\Lambda$CDM and is with in the error-bars of
Planck 2018 data~\cite{Aghanim:2019ame}. In the low-$\ell$, we can see
the CMB value of GP Theory is smaller than $\Lambda$CDM when $\ell$ is
smaller to 10. This difference in large scale structure shows the
impact of the GP Theory's initial condition in the early universe, but
it is still under the error-bar and hard to be distinguished in
present CMB observational data as shown in the figure. The lower panel
of the Fig.~\ref{fig:cmbTT} shows the relative difference of the
$C_{\ell}^{TT}$ of GP theory with respect to $\Lambda$CDM. The
difference is inside the error-bars.

\section{Conclusion\label{conclu}}

In this work we analyze extensively the viability of the generalized
Proca theory up to cubic order in the Lagrangian from the cosmological
observation. This study is particularly interesting in the context of
the present day $H_{0}$ tension that is getting stronger. On
considering Planck data and $H_{0}$ measurements alone, within this
theory we do not see any tension between CMB data and the local
measurement of the current expansion rate of the universe. In fact, we
find a mean value for $H_{0}$ which exactly matches the value measured
by $\text{H}0\text{LiCOW}$, namely $H_{0}=$$73.34_{-2.69}^{+2.46}$ at
$95\%$ C.L. However, using BAO, HST, Planck and JLA together we see
that the best fit for $H_{0}$ tends to be reduced, but still $H_{0}$
measurements are inside the 2 sigma contours. This indicates that the
GP theory is able to reduce the tension below 2 sigma.

For this theory there are three more parameters
$s,\,p_{2},\,\bar{\varphi}_{0}$ in comparison with standard
$\Lambda\text{CDM}$ cosmological model.  Out of which only $s$ affects
the background dynamics. In particular, the parameter $s$ defines how
much the theory deviates from $\Lambda\text{CDM}$ (which is obtained
for the background in the limit $s\to0$). In other words in the limit
$s\rightarrow0$ at the background level the theory becomes
$\Lambda\text{CDM}$.

From the Planck + HST the parameter $s$ is constrained to be
$s=0.334_{-0.211}^{+0.358}$ at $95\%$ C.L.\, and on adding JLA and BAO
we get $s=0.172_{-0.029}^{+0.026}$ at $95\%$ C.L.. The value of $s$ is
in agreement with the previous study of the same theory with the CMB
shift parameter~\cite{deFelice:2017paw}.  However, our results are not
only updated to the latest Planck and $H_{0}$ results, but the
methodology is quite different. In~\cite{deFelice:2017paw}, Planck
data were considered only through the constraints on the CMB-shift
parameters. Instead in this study of ours, we have used the whole
Planck data at once. Therefore, we can say we have confirmed the
behaviour but such a step was a non-trivial one to show. In fact,
other cases are known where the results from these two different
methodologies do not agree with each other (see
e.g.\ \cite{Dirian:2014bma}).  The situation is also similar for the
parameter $p_{2}$ at the $95\%$ C.L.\ for Planck+HST is
$p_{2}=0.8344_{-0.6459}^{+0.4390}$ and with other data sets
$p_{2}=0.226_{-0.087}^{+0.099}$. While the parameter
$\bar{\varphi}_{0}$ does not converge at the $95\%$ C.L. there is
large degeneracy for both cases. This is clearly shown in the
Fig.\ \ref{fig:pla+hst_tri}.

Another remarkable point that has to be emphasized is that for the
both MCMC run that is Planck+HST and Planck+HST+BAO+JLA the GP theory
shows better fits in comparison with $\Lambda\text{CDM}$. The
difference in $\chi^{2}$ is $|\Delta\chi^{2}|=22$ and
$|\Delta\chi^{2}|=7$ respectively for Planck+HST and
Planck+HST+BAO+JLA. This kind of behaviour is seen in other modified
theories of gravity, for example~\cite{Frusciante:2019puu} (also
see~\cite{Li:2020ybr,Li:2019ypi}). Nonetheless, we are able to show
that Planck data and $H_{0}$ measurements agree with each other at
2-sigma within the GP theory. In future further exploring the reason
behind the preference of models other than $\Lambda\text{CDM}$ will be
of particular interest.

This study once again shows that the generalized proca theory up to
the cubic order terms in the Lagrangian (as to have a speed of
propagation for the gravitational wave $c_{T}\equiv1$) reduces the
tension of $H_{0}$ below 2-sigma in present data sets. This is mostly
due to the particular background dynamics of
$\Omega_{\text{DE}}$. This solution can be thought of as post
recombination approach to the $H_{0}$ tension with modified gravity.

\acknowledgments

ADF acknowledges support from National Center for Theoretical Sciences
during his stay in Taiwan. ADF wants to thank prof.\ Maggiore for the
encouragement to start the study of Boltzmann code
solvers. M.~C.~P.~acknowledges the support from the Japanese
Government (MEXT) scholarship for Research Student. CQG and LY were
supported in part by National Center for Theoretical Sciences and MoST
(MoST-107-2119-M-007-013-MY3). The numerical computation in this work
was carried out at the Yukawa Institute Computer Facility.

\appendix

\section{Linear perturbation equations of motion \label{pert_eoms}}

For clarity, here we give the full perturbation equations of
motion. These same equation can be found in the C-code we have
provided. After the discussion of section~\ref{pert}, the modified
equations of motion for the perturbations can be written as follows
\begin{eqnarray}
  \ddot{\bar J}&=&\left( 3/2\,{\frac {{\bar\varphi}\,{a}^{3} \left(
      p+1 \right) {\Xi_{\rm T}}}{{p}^ {2}{{\mathcal H}}^{2} \left(
      {\Omega_{\rm DE}}\,s+1 \right) }}-2\,{\frac {{\Omega_{\rm
          DE}}\,as}{ {\bar\varphi}}} \right) {\dot\phi}-{\frac
    {{\bar\varphi}\,a \left( p+1 \right)
      {\dot\phi}}{p}}\nonumber\\ &+& \left( -3\,{\frac { \left( ps-1
      \right) {a}^{2}{k}^{2 }{{\bar\varphi}}^{2}{\Xi_{\rm
          T}}}{p{\mathcal H} \left( 6\,{{\mathcal
          H}}^{2}{p}^{2}s{\Omega_{\rm DE}}+{k}^{2}{ {\bar\varphi}}^{2}
      \right) \left( {\Omega_{\rm DE}}\,s+1 \right) }}-2\,{\frac
    {{k}^{2} {\mathcal H}{{\bar\varphi}}^{2}}{6\,{{\mathcal
          H}}^{2}{p}^{2}s{\Omega_{\rm
          DE}}+{k}^{2}{{\bar\varphi}}^{2}}} \right) {\dot{\bar
      J}}\nonumber\\ &+& \left( -{\frac {{a}^{2} \left( {k}^{4} \left(
      -1/2+ \left( s+1/2 \right) p \right)
      {{\bar\varphi}}^{4}+3\,{k}^{2}{\Omega_{\rm DE}}\,{p}^{2
      }s{{\mathcal H}}^{2} \left( p+2 \right)
      {{\bar\varphi}}^{2}+18\,{{\Omega_{\rm DE}}}^{2}{p}^{4}{s}^
      {2}{{\mathcal H}}^{4} \right) {\Xi_{\rm T}}}{{p}^{2}{{\mathcal
          H}}^{2}{{\bar\varphi}}^{2} \left( 6\,{{\mathcal H}}^
      {2}{p}^{2}s{\Omega_{\rm DE}}+{k}^{2}{{\bar\varphi}}^{2} \right)
      \left( {\Omega_{\rm DE}}\,s+1 \right)
  }}\right.\nonumber\\ &-&\left.2/3\,{\frac { \left(
      1/2\,{k}^{2}{{\bar\varphi}}^{4}+{\Omega_{\rm DE}}\,ps \left(
      -3\,{{\mathcal H}}^{2}p+{k}^{2} \right)
      {{\bar\varphi}}^{2}+6\,{{\Omega_{\rm DE}}}^{2}{p}^
      {3}{s}^{2}{{\mathcal H}}^{2} \right)
      {k}^{2}}{p{{\bar\varphi}}^{2} \left( 6\,{{\mathcal H}}^{2}{p}
      ^{2}s{\Omega_{\rm DE}}+{k}^{2}{{\bar\varphi}}^{2} \right) }}
  \right) {\bar J}\nonumber\\ &+& \left( -3\,{ \frac
    {{\bar\varphi}\,{a}^{3} \left( p+1 \right) \left(
      {k}^{2}{{\bar\varphi}}^{2
      }ps-1/2\,{k}^{2}{{\bar\varphi}}^{2}+3\,{{\mathcal
          H}}^{2}{p}^{2}s{\Omega_{\rm DE}} \right) {\Xi_{\rm T}}
    }{{p}^{2}{\mathcal H} \left( 6\,{{\mathcal
          H}}^{2}{p}^{2}s{\Omega_{\rm DE}}+{k}^{2}{{\bar\varphi}}^{2}
      \right) \left( {\Omega_{\rm DE}}\,s+1 \right)
  }}\right.\nonumber\\ &+&\left. {\frac {a \left( -36\,{p}
      ^{2}{\Omega_{\rm DE}}\, \left( \left( p/2+1/2 \right)
      {{\bar\varphi}}^{2}+ps{\Omega_{\rm DE}} \right) s{{\mathcal
          H}}^{4}+6\,{{\bar\varphi}}^{2} \left( \left( -3/2\,p-3/2
      \right) {{\bar\varphi}}^{2}+ps{\Omega_{\rm DE}}\, \left( p-1
      \right) \right) {k}^{2}{{\mathcal H}}^{2}+{k
      }^{4}{{\bar\varphi}}^{4} \right) }{3p{\mathcal H}{\bar\varphi}\,
      \left( 6\,{{\mathcal H}}^{2}{p}^{2}s {\Omega_{\rm
          DE}}+{k}^{2}{{\bar\varphi}}^{2} \right) }} \right)
  \psi\nonumber\\ &+& \left( \frac92{ \frac { ( p+1 )
      {\bar\varphi}{a}^{5} \left( 3{{\mathcal
          H}}^{2}{p}^{2}s{\Omega_{\rm DE}}\, ( {\Omega_{\rm DE}}s+1 )
      -6s{\Omega_{\rm DE}}{{\mathcal H}}^{2 } ( {\Omega_{\rm
          DE}}{s}^{2}-1 ) {p}^{3}+{k}^{2}{{\bar\varphi}}^{2} ( s+1 ) p
      -1/2\,{k}^{2}{{\bar\varphi}}^{2} \left( {\Omega_{\rm DE}}\,s+1
      \right) \right) { {\Xi_{\rm T}^2}}}{ \left( {\Omega_{\rm
          DE}}\,s+1 \right) ^{3} \left( 6\,{{\mathcal H}}^{2}{p}^{2
      }s{\Omega_{\rm DE}}+{k}^{2}{{\bar\varphi}}^{2} \right)
      {{\mathcal H}}^{3}{p}^{3}}}\right.\nonumber\\ &+&{\frac {{a}^{3}
      \left( {k}^{4} \left( -1/2+ \left( s+1/2 \right) p \right)
           {{\bar\varphi}}^ {4}+3\, \left( -6\,{p}^{2}{{\mathcal
               H}}^{2}+ \left( -6\,{{\mathcal H}}^{2}+{k}^{2} \right)
           p+2\,{k}^{2} \right) {p}^{2}{\Omega_{\rm DE}}\,{{\mathcal
               H}}^{2}s{{\bar\varphi}}^{2}+18\,{{\Omega_{\rm DE}}}
           ^{2}{p}^{4}{s}^{2}{{\mathcal H}}^{4} \right) {\Xi_{\rm
               T}}}{{\bar\varphi}\,{p}^{3}{{\mathcal H}}^{3} \left(
      6\,{{\mathcal H}}^{2}{p}^{2}s{\Omega_{\rm
          DE}}+{k}^{2}{{\bar\varphi}}^{2} \right) \left( {\Omega_{\rm
          DE}}\,s+1 \right) }}\nonumber\\ &-&\left.2\,{\frac
    {{\bar\varphi}\,{a}^{3} \left( p+1 \right)
      {\Xi_r}}{{p}^{2}{\mathcal H} \left( {\Omega_{\rm DE}}\,s+1
      \right) }}+2/3\,{\frac {a \left( 1/2
      \,{k}^{2}{{\bar\varphi}}^{4}+{\Omega_{\rm DE}}\,ps \left(
      -3\,{{\mathcal H}}^{2}p+{k}^{2} \right)
      {{\bar\varphi}}^{2}+6\,{{\Omega_{\rm
            DE}}}^{2}{p}^{3}{s}^{2}{{\mathcal H}}^{2} \right)
      {k}^{2}}{{p }^{2}{\mathcal H}{\bar\varphi}\, \left(
      6\,{{\mathcal H}}^{2}{p}^{2}s{\Omega_{\rm
          DE}}+{k}^{2}{{\bar\varphi}}^{2} \right) }} \right) \phi
  \,,\\ \dot{\phi}&=&-1/4\,{\frac { \left( -4/3\,{{\Omega_{\rm
            DE}}}^{2}{p}^{2}{s}^{2}{{\mathcal H}}^{2}-4/3\,{{\mathcal
          H}} ^{2}{p}^{2}s{\Omega_{\rm
          DE}}+{a}^{2}{{\bar\varphi}}^{2}{\Xi_{\rm T}}\, \left( p+1
      \right) \right) s{\Omega_{\rm DE}}\,{k}^{2}\phi}{{\mathcal H}
      \left( {\Omega_{\rm DE}}\,s+1 \right) ^{2}p \left( {{\mathcal
          H}}^{2}{p}^{2}s{\Omega_{\rm
          DE}}+1/6\,{k}^{2}{{\bar\varphi}}^{2} \right)
  }}\nonumber\\ &-&{ \frac { \left( 6\,{{\Omega_{\rm
            DE}}}^{2}{p}^{2}{s}^{2}{{\mathcal H}}^{2}-{\Omega_{\rm
          DE}}\,{k}^{2}ps{ {\bar\varphi}}^{2}+6\,{{\mathcal
          H}}^{2}{p}^{2}s{\Omega_{\rm DE}}+{k}^{2}{{\bar\varphi}}^{2}
      \right) {\mathcal H} \psi}{ \left( {\Omega_{\rm DE}}\,s+1
      \right) \left( 6\,{{\mathcal H}}^{2}{p}^{2}s{\Omega_{\rm
          DE}}+{k} ^{2}{{\bar\varphi}}^{2} \right)
  }}\nonumber\\ &+&{\frac {{k}^{2}{\Omega_{\rm DE}}\,ps{\mathcal
        H}{\bar\varphi}\,{\dot{\bar J}}}{a \left( {\Omega_{\rm
          DE}}\,s+1 \right) \left( 6\,{{\mathcal
          H}}^{2}{p}^{2}s{\Omega_{\rm DE}}+{k}^{ 2}{{\bar\varphi}}^{2}
      \right) }}-2\,{\frac {{{\Omega_{\rm
            DE}}}^{2}{k}^{2}{p}^{2}{s}^{2 }{{\mathcal H}}^{2}{\bar
        J}}{{\bar\varphi}\,a \left( {\Omega_{\rm DE}}\,s+1 \right)
      \left( 6\,{{\mathcal H}}^{2}{p} ^{2}s{\Omega_{\rm
          DE}}+{k}^{2}{{\bar\varphi}}^{2} \right) }}+\frac32\,{\frac
    {{a}^{2} {\sum_i[(\varrho_i+p_i)\theta_i]}}{{k}^{2} \left(
                    {\Omega_{\rm DE}}\,s+1 \right) }}\,,\\
  \Xi_{\rm T}&=&\sum_i(\varrho_i+p_i)\,,\qquad \Xi_r = \sum_j\varrho_j\,,\qquad \mathcal{H} = \frac{\dot{a}}{a}\,,
\end{eqnarray}
whereas the field $\psi$ is given by exactly the same shear equation
of General Relativity (see Eq.\ (\ref{eq:shearGR})). Furthermore the
sum over $i$ runs over the standard matter species, whereas $j$ only
over the radiation-like ones (this term comes from the presence of
$\dot p$ terms, to which dust does not give any contribution). Let us
also remind ourselves about the definitions of $\varrho$ and
$p$ given in Eqs.\ (\ref{eq:rhoCL}) and~(\ref{eq:pCL}).
 
\bibliographystyle{unsrt} \bibliography{bibliography}

\end{document}